\documentclass{jpsj-suppl}
\usepackage{txfonts} 
\usepackage[dvipdfmx]{graphicx}

\title{Microscopic Calculations of Vortex-Nucleus Interaction in the Neutron Star Crust}

\author{
Kazuyuki \textsc{Sekizawa}$^{1}$, Gabriel \textsc{Wlaz{\l}owski}$^{1,2}$, Piotr \textsc{Magierski}$^{1,2}$,
Aurel \textsc{Bulgac}$^{2}$, and Michael McNeil \textsc{Forbes}$^{2,3}$
}

\inst{
$^{1}$Faculty of Physics, Warsaw University of Technology, ulica Koszykowa 75, 00-662 Warsaw, Poland\\
$^{2}$Department of Physics, University of Washington, Seattle, WA 98195-1560, USA\\
$^{3}$Department of Physics \& Astronomy, Washington State University, Pullman, WA 98164-2814, USA
}

\email{sekizawa@if.pw.edu.pl}

\recdate{September 11, 2016}

\abst{
We investigate the dynamics of a quantized vortex and a nuclear impurity
immersed in a neutron superfluid within a fully microscopic time-dependent 
three-dimensional approach. The magnitude and even the sign of the force 
between the quantized vortex and the nuclear impurity have been a matter
of debate for over four decades. We determine that the vortex and the impurity
repel at neutron densities, $0.014$~fm$^{-3}$ and $0.031$~fm$^{-3}$,
which are relevant to the neutron star crust and the origin of glitches, while
previous calculations have concluded that the force changes its sign between
these two densities and predicted contradictory signs. The magnitude of the
force increases with the density of neutron superfluid, while the magnitude
of the pairing gap decreases in this density range.
}

\kword{pulsar glitch, vortex pinning, superfluid, inner crust, neutron star}

\begin{document}

\begin{flushright}
NT@UW-16-11 
\vspace{-5mm}
\end{flushright}

\maketitle

\section{Introduction}

The origin of a neutron star glitch, a sudden spin-up of the rotational frequency,
has been one of the unsolved problems in nuclear astrophysics for a long time.
It has been suggested \cite{Anderson-Itoh} that the glitch is caused by a catastrophic
unpinning of a huge number of vortices from the pinning sites formed by a Coulomb
lattice of nuclei immersed in a neutron superfluid in the inner crust of neutron stars.
Although the vortex-nucleus interaction is undoubtedly one of the most important
ingredients needed to explain the glitches, so far contradictory predictions have been
made about both its magnitude and even sign. The difficulty in extracting the vortex-nucleus
interaction is related to the enormous number of degrees of freedom that one has to
take into account. The vortex itself represents a topological excitation of a superfluid,
which may stretch and bend in various ways and the susceptibility towards vortex
deformations should be derived from a microscopic equation of motion describing
neutron superfluid. Moreover, the nuclear impurity may easily deform as its surface
tension is significantly smaller than those of isolated nuclei. Thus various assumptions
concerning the symmetry of the problem considered in the past, in order to simplify
the analyses, may dramatically change results, as one can accidentally omit important
degrees of freedom of either the vortex or of the impurity.
 
Recently, we have reported the first fully microscopic, three-dimensional and
symmetry-unconstrained dynamical simulations which enabled us  to extract
the vortex-nucleus interaction~\cite{vortex}.

\section{Methods}

\subsection{TDSLDA}

We have used an extension of Kohn-Sham density functional theory
for superfluid systems, known as the time-dependent superfluid local
density approximation (TDSLDA). It has been proved that the TDSLDA
provides very accurate description for the static and dynamic properties
of strongly correlated fermionic systems such as ultracold atomic gases
\cite{coldatoms1} and nuclear systems \cite{nuclei1}.

For the normal part of the functional, we used the FaNDF$^0$ functional
constructed by Fayans \textit{et al.} \cite{Fayans1}, which reproduces
the infinite matter equation of state of Refs.~\cite{EOS1,EOS2} and various
properties of finite nuclei \cite{nuclear1,nuclear2}. In this study we have omitted
the spin-orbit term, as it plays a minor role in the vortex-nucleus dynamics \cite{vortex}.
To the FaNDF$^0$ functional we added a contribution describing the pairing correlations,
$\mathcal{E}_{\rm pair}(\boldsymbol{r})=g(n_n(\boldsymbol{r}))|\nu_n(\boldsymbol{r})|^2+g(n_p(\boldsymbol{r}))
|\nu_p(\boldsymbol{r})|^2$, where $\nu_{n,p}$ are the anomalous densities of
neutrons and protons. The coupling constant $g$ depends on densities of neutrons
$n_n$ and protons $n_p$, which was adjusted to reproduce the BCS-type neutron
pairing gap in pure neutron matter~\cite{vortex}.

\subsection{Computational settings}

We solved TDSLDA equations (formally equivalent to the time-dependent
Hartree-Fock-Bogoliubov or time-dependent Bogoliubov-de Gennes equations)
on a periodic 3D lattice without any symmetry restrictions. We used a box of
$75\;{\rm fm}\,\times75\;{\rm fm}\,\times\,60\;{\rm fm}$ with a lattice spacing
of 1.5~fm, which corresponds to a rather large momentum cutoff, $p_{\rm c} \approx
400$~MeV/$c$. To prevent a quantum vortex from interacting with vortices in neighboring
cells, due to periodic boundary conditions, we introduced a flat-bottomed external potential
confining the system in a tube along $z$-axis with a radius 30~fm. Spatial derivatives
are evaluated using Fourier transformation. We used the Trotter-Suzuki decomposition
for the time evolution operator with a predictor-corrector step with a time-step set to
$\Delta t=0.054$~fm/$c$. We generated initial states for two background neutron
densities, $n=0.014$~fm$^{-3}$ and $0.031$~fm$^{-3}$, with a vortex line
parallel to the axis of the tube and with a nuclear impurity with 50 protons \cite{COCGSupp}.

\section{Force extraction from dynamical simulations}

We extracted the force exerted on the nucleus from our dynamical simulations.
We performed the time integration with an external force $\boldsymbol{F}_{\rm ext}(t)$
constant in space, which couples to protons only. This force moves the
center-of-mass of protons together with bound neutrons in the nucleus.
Naturally, unbound neutrons in the vicinity of the nucleus are also entrained.
Assuming that there are only two forces acting on the nucleus, $\boldsymbol{F}_{\rm ext}(t)$
and $\boldsymbol{F}(t)$, where the latter is the force arising from the interaction
with the vortex, the nucleus will move with constant velocity if the relation
$\boldsymbol{F}(t)+\boldsymbol{F}_{\rm ext}(t)=0$ holds. Thus by adjusting
$\boldsymbol{F}_{\rm ext}(t)$ we may extract the force $\boldsymbol{F}(t)$
exerted on the nucleus by the vortex. In practice we adjusted the external force
during time evolution as
\begin{equation}
\boldsymbol{F}_{\rm ext}(t+\Delta t)= \boldsymbol{F}_{\rm ext}(t) - \alpha[\boldsymbol{v}(t)-\boldsymbol{v}_0],
\end{equation}
where $\boldsymbol{v}(t)$ is the velocity of the center-of-mass of protons and
$\alpha$ is a small coefficient governing the rate of the force adjustment. In our
simulations, we dragged the nucleus with a very small velocity, $v_0=0.001c$,
which is far below the critical velocity to ensure that no phonons are excited.

\begin{figure}[t]
   \begin{center}
   \includegraphics[width=\columnwidth]{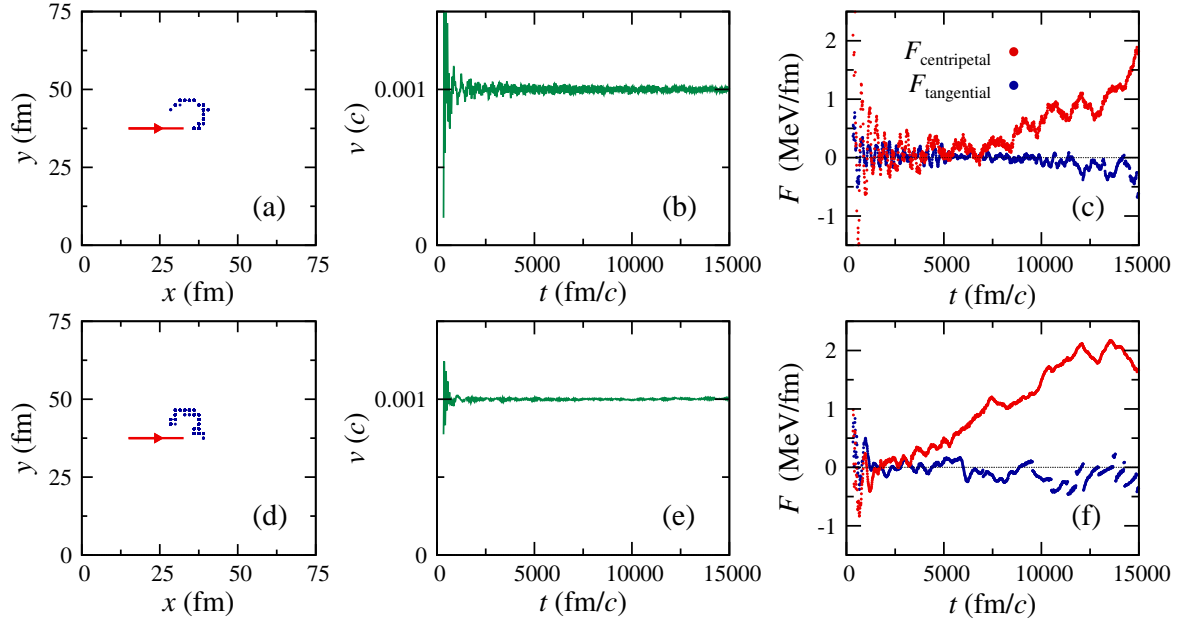}
   \end{center}\vspace{-2.6mm}
   \caption{
   The results of the TDSLDA calculations for $n=0.014$~fm$^{-3}$ (a,~b,~c)
   and $0.031$~fm$^{-3}$ (d,~e,~f). The trajectory of the motion of the vortex
   core is displayed in subfigures (a,~d). The position of the vortex core is shown in
   the $xy$-plane, where the nucleus resides. The red line shows the trajectory of
   the nucleus and the blue dots show the vortex-core positions at various times.
   The nucleus velocity is shown as a function of time in subfigures (b,~e). The
   central (red dots) and the tangential (blue dots) components of the extracted
   force $\boldsymbol{F}$ are shown as a function of time in (c,~f).
   }\vspace{-5mm}
   \label{FIG:results}
\end{figure}

To extract the vortex-nucleus force as a function of their separation $R$,
we started simulations from a well separated configuration and dragged the
nucleus along the positive-$x$ direction towards the vortex. We displayed the
results of the TDSLDA calculations for $n=0.014$~fm$^{-3}$ in Fig.~\ref{FIG:results}~(a,~b,~c)
and for $n=0.031$~fm$^{-3}$ in Fig.~\ref{FIG:results}~(d,~e,~f), respectively.

Figure~\ref{FIG:results}~(a,~d) illustrates a typical trajectory of the motion
of the vortex core in the $xy$ plane, where the center-of-mass of the nucleus
resides ($z=30$~fm). The red arrow indicates the trajectory of the center-of-mass
of the nucleus. The blue dots show the positions of the vortex core at various times.
As the nucleus approaches the vortex, it exerts a repulsive force $\boldsymbol{F}$
on the vortex. The vortex responds it, according to the formula $\boldsymbol{F}_{\rm M}
\propto \boldsymbol{\kappa}\times\boldsymbol{v}$, where $\boldsymbol{\kappa}$
is the circulation pointing in the positive-$z$ direction (upward perpendicular to the figure). For both densities, we found
that the vortex, under the perturbation induced by the nucleus, is initially moving
along the positive-$y$ direction. It clearly indicates that the force is repulsive and
initially directed along the positive-$x$ direction away from the nucleus.

In Fig.~\ref{FIG:results}~(b,~e), we show the magnitude of the velocity of the
nucleus as a function of time. The velocity is adjusted during the first few 1,000~fm/$c$,
and subsequently is kept constant during the time evolution. It allows us to extract
the force $\boldsymbol{F}(t)$ exerted on the nucleus as discussed above.

In Fig.~\ref{FIG:results}~(c,~f), we show the force $\boldsymbol{F}$ as a
function of time. We decomposed the force into central and tangential components
with respect to the vortex-core position at each time. The extracted force is, as
expected, predominantly central with a negligibly small tangential component.
The magnitude of the force becomes stronger for higher density. The effective
range of the force increases with density, consistent with a smaller pairing gap
and a larger coherence length at higher densities.

\begin{figure}[t]
   \begin{center}
   \includegraphics[width=\columnwidth]{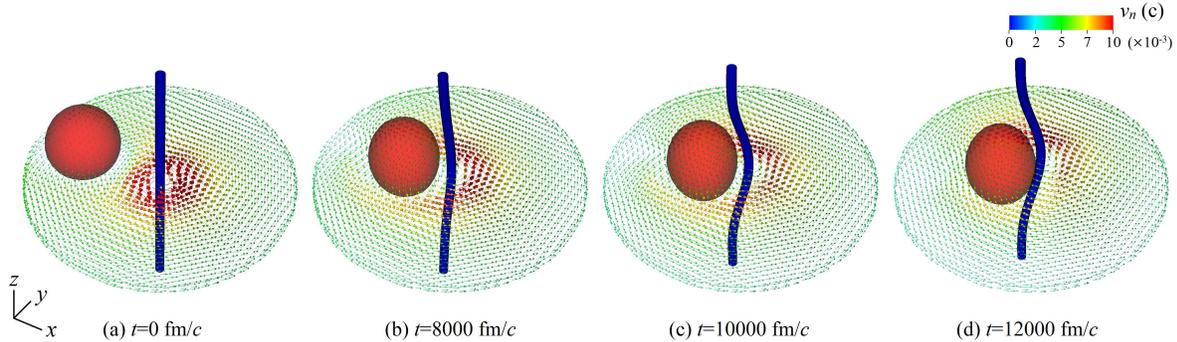}
   \end{center}\vspace{-1.4mm}
   \caption{
   Snapshots of the system ($n=0.031$~fm$^{-3}$) for times
   $(0, 8, 10, 12)\times 10^3$~fm/$c$. The blue tube represents position
   of the vortex core. The red contour indicates the surface of the nuclear
   impurity (where proton density drops to value $0.01$~fm$^{-3}$).
   Velocity field of neutrons is presented by vectors (color indicates its magnitude).
   }\vspace{-4mm}
   \label{FIG:shape}
\end{figure}

Figure~\ref{FIG:shape} displays snapshots of the system ($n=0.031$~fm$^{-3}$)
exhibiting shape of the nucleus and the vortex line as well as the velocity field
of neutrons. While the velocity field inside the nucleus is smaller than outside
(about $0.005c$), which is about 5 times larger than the dragging velocity.
Thus the dragging process does not influence strongly the superflow of neutrons.
As the structure of both objects is affected by their interaction (the vortex bends
and the nucleus deforms), the force cannot be described by a simple function of
their separation only. In Ref.~\cite{vortex}, we have deduced force per unit length
to properly characterize the force for various vortex-nucleus configurations.

\section{Summary}

We have performed 3D, symmetry unrestricted, microscopic, dynamic
simulations for a vortex-nucleus system using a time-dependent extension
of DFT for superfluid systems. We have determined that the vortex-nucleus
force is repulsive and increasing in magnitude with density, for the densities
characteristic of the neutron star crust (0.014~fm$^{-3}$ and 0.031~fm$^{-3}$). 

It is instructive to note that repulsive force is not ruled out by a purely
hydrodynamical approach (see Supplemental Material of \cite{vortex}).
It is however difficult to associate unambiguously the superfluid density with
the actual neutron density. Therefore hydrodynamical description can be
used only to estimate the asymptotic behavior of the vortex-nucleus interaction
($\propto 1/r^3$).

It is worth mentioning that the extracted force is at least one order of
magnitude larger than those predicted in a recent phenomenological
analysis~\cite{Seveso(2016)}. The repulsive force is compatible with
the so-called interstitial pinning, where vortices are trapped at positions
that maximize the overall separation from the nearest nuclei \cite{Link(2009)}.

\section*{Acknowledgments}

We are grateful to K.J. Roche and A. Sedrakian for helpful discussions
and comments. This work was supported by the Polish National Science
Center (NCN) under Contracts No. UMO-2013/08/A/ST3/00708 and
No. UMO-2014/13/D/ST3/01940. Calculations have been performed:
at the OLCF Titan---resources of the Oak Ridge Leadership Computing
Facility, which is a DOE Office of Science User Facility supported under
Contract DE-AC05-00OR22725; at NERSC Edison---resources of the
National Energy Research Scientiffic computing Center, which is supported
by the Office of Science of the U.S. Department of Energy under Contract
No. DE-AC02-05CH11231; at HA-PACS system---resources provided by
Interdisciplinary Computational Science Program in Center for Computational
Sciences, University of Tsukuba. This work was supported in part by U.S.
DOE Office of Science Grant DE-FG02-97ER41014, and a WSU New Faculty
Seed Grant.

\end{document}